\documentclass[12pt]{article}
\usepackage{amsmath}
\usepackage{epsf}
\usepackage{epsfig}
\usepackage{here}
\usepackage{amssymb}
\usepackage{citesort}
\usepackage{graphicx}
\usepackage{latexsym}
\newcommand{\be}{\begin{equation}}
\newcommand{\ee}{\end{equation}}
\newcommand{\bear}{\begin{eqnarray}}
\newcommand{\ear}{\end{eqnarray}}

\newsavebox{\LSIM}
\sbox{\LSIM}{\raisebox{-1ex}{$\ \stackrel{\textstyle<}{\sim}\ $}}
\newcommand{\lsim}{\usebox{\LSIM}}
\newsavebox{\GSIM}
\sbox{\GSIM}{\raisebox{-1ex}{$\ \stackrel{\textstyle>}{\sim}\ $}}
\newcommand{\gsim}{\usebox{\GSIM}}

\begin{document}
\begin{titlepage}
\begin{flushright}
BA-01-19\\
hep-ph/0104293
\end{flushright}
$\mbox{ }$
\vspace{.1cm}
\begin{center}
\vspace{.5cm}
{\bf\Large Neutrino oscillations and rare processes}\\[.3cm]
{\bf\Large in models with a small extra dimension}\\
\vspace{1cm}
Stephan J. Huber\footnote{shuber@bartol.udel.edu} 
and
Qaisar Shafi\footnote{shafi@bartol.udel.edu} \\

\vspace{1cm} {\em 
Bartol Research Institute, University of Delaware, Newark, DE 19716, USA
}
\end{center}
\bigskip\noindent
\vspace{1.cm}
\begin{abstract}
 We discuss Dirac neutrino masses and mixings in a scenario where both the
standard model fermions and right handed neutrinos are bulk fields in
a non-factorizable geometry in five dimensions. We show how the atmospheric
and solar neutrino anomalies can be satisfactorily resolved, and in particular
how bimaximal mixing is realized. We also consider rare processes such as
neutron-antineutron oscillations and $\mu\rightarrow e + \gamma$, 
which may occur at an observable rate. 
\end{abstract}
\end{titlepage}
\section{Introduction}
The huge discrepancy between the Planck scale 
$M_{\rm Pl}\sim10^{19}$ GeV and the scale of
electroweak symmetry breaking $M_Z\sim10^{2}$ GeV,
is one of the most interesting challenges in modern
physics. Recently it was realized that compact extra 
space dimensions can offer a new perspective on this 
gauge hierarchy problem. In one class of models the 
weakness of 4-dimensional gravity is induced by the 
very large volume of compactification \cite{ADD}.  
Subsequently, it was demonstrated
that a small but warped extra dimension provides
an elegant alternative solution to the hierarchy 
problem \cite{RS} (see also \cite{G}). The fifth 
dimension is an $S_1/Z_2$
orbifold with an AdS$_5$ geometry, bordered
by two 3-branes with opposite tensions and separated
by distance $R$.  The 
AdS warp factor $\Omega=e^{-\pi k R}$
generates an exponential hierarchy between the
effective mass scales on the two branes ($k$=AdS curvature).
If the brane separation is $kR\simeq 11$, the scale on 
the negative tension brane is of TeV-size, while the scale on 
the other brane is of order $M_{\rm Pl}$.  The 
AdS curvature $k$ and the 5d Planck mass $M_5$ 
are both assumed to be of order $M_{\rm Pl}$. At the 
TeV-brane gravity is weak because the zero mode
corresponding to the 4d graviton is localized at the
positive tension brane (Planck-brane). 

In low scale models of quantum gravity higher-dimensional
operators, now only TeV-scale suppressed, are known to induce 
rapid proton decay, large neutrino masses and flavor violating 
interactions. In models with non-factorizable (warped) geometry these problems 
can be cured to some extent by introducing the Standard Model (SM)
fermions as bulk fields, without relying on ad-hoc symmetries such
as lepton or baryon number \cite{GP,HS2}. Because of the 
warp factor, the effective cut-off scale varies along the extra
dimension. If the quarks and leptons are localized towards
the Planck-brane in the extra dimension, the effective cut-off scale 
can by much larger than a TeV. 

Fermion masses are generated by the Higgs mechanism.
In the non-supersymmetric framework we are discussing,
the Higgs field lives close to the TeV-brane in order 
to maintain the solution of the gauge hierarchy problem 
\cite{CHNOY,HS,DHR}. The induced fermion masses crucially
depend on the overlap between the Higgs and fermion wave
functions in the extra dimension, and naturally become small
for fermions residing close to the Planck-brane.
This mechanism offers a higher-dimensional view on the problem
of fermion mass hierarchies \cite{GP}. Also, the quark mixings
can be nicely reproduced from order unity Yukawa couplings \cite{HS2}.
The nearest neighbor structure in the CKM matrix is reflected in the
different positions the fermions occupy in the extra dimension.

In our set-up we assume that the SM gauge bosons
are bulk fields as well in order to guarantee bulk gauge invariance.
Electroweak precision data \cite{DHRP,CHNOY,GP,DHR}, especially
the W and Z boson mass ratio, impose stringent constraints by
requiring the Kaluza Klein (KK) excitations of bulk gauge bosons and
fermions to have masses of order 10 TeV \cite{HS}. 
The measured W and Z boson masses are reproduced
by some (mild) tuning of parameters. We note in passing that
contributions of 10 TeV Kaluza-Klein states to the anomalous magnetic 
moment of the muon are two to three orders of magnitude
smaller \cite{DHR2} than the recently reported deviation of 
this observable from its SM prediction \cite{g_2}. 
  
Taking into account the constraints on the locations of 
fermions from their masses and mixings, one can more reliably
estimate the suppression scales for higher-dimensional
operators. Even though flavor violating interactions are
safely suppressed, four-fermion operators suppressed by mass scales of
about $10^{12}$ GeV can still lead to proton decay at an unacceptable
level \cite{HS2}, unless the dimensionless couplings happen to be
$\lsim 10^{-7}$. In addition keV-size Majorana neutrino masses
and small mixings are induced, which do not easily fit the neutrino 
oscillation data \cite{osc}. 

In this letter we study the generation of Dirac masses for the
neutrinos by introducing right-handed (sterile) neutrinos in the
bulk. In the case of TeV-brane SM neutrinos this scenario was 
put forward in ref.~\cite{GN}. It turns out that we can accommodate 
the atmospheric and solar neutrino oscillations without fine-tuning 
of parameters. Having the SM neutrinos in the bulk 
reduces their mixing with Kaluza-Klein sterile neutrinos. Lepton
flavor violating processes, which in the scenario of ref.~\cite{GN}
are in conflict with experiment for Kaluza Klein masses below 
25 TeV \cite{K00}, are safely suppressed. In this respect, moving
the SM fermions off the TeV-brane  allows for smaller KK masses $\sim 10$ TeV
and therefore reduces the amount  of
fine-tuning to obtain realistic weak boson masses. 
The unacceptably large Majorana neutrino masses 
(see previous paragraph) are eliminated by imposing
lepton number symmetry. Another important consequence is that 
the proton becomes stable. However, certain 
baryon number violating processes, e.g.~neutron anti-neutron oscillations
with $\Delta B=2$, are still possible and may lead to interesting
experimental signatures \cite{BD,CC01}.  

%
%
%
%%%%%%%%%%%%%%%%%%%%%%%%%%%%%%%%%%%%%%%%%%%%%%%%%%%%%%%%%%%%%%%%%%%%%%%%%%%%%%%%%
%
%
%
\section{Bulk fermions}
To set the notation let us briefly review some properties of
fermions in a slice of AdS$_5$.
We consider the non-factorizable metric \cite{RS}
\begin{equation} \label{met}
ds^2=e^{-2\sigma(y)}\eta_{\mu\nu}dx^{\mu}dx^{\nu}+dy^2,
\end{equation}
where $\sigma(y)=k|y|$.  
The 4-dimensional metric is $\eta_{\mu\nu}={\rm diag}(-1,1,1,1)$, 
$k$ is the AdS curvature related to the bulk cosmological constant
and brane tensions, and $y$ denotes the fifth coordinate.
The equation of motion for a fermion in curved space-time reads
\begin{equation}
E_a^M\gamma^a(\partial_M+\omega_M)\Psi+m_{\Psi}\Psi=0,
\end{equation}
where $E_a^M$ is the f\"unfbein, $\gamma^a=(\gamma^{\mu},\gamma^5)$ 
are the Dirac matrices in flat space, 
\begin{equation}
\omega_M=\left(\frac{1}{2}e^{-\sigma}\sigma'\gamma_5\gamma_{\mu},0\right)
\end{equation}
 is the spin connection, and $\sigma'=d\sigma/dy$.
The index $M$ refers to objects in 5d curved space,
the index $a$ to those in tangent space. 
Fermions have two possible transformation properties 
under the $Z_2$ orbifold symmetry,
$\Psi(-y)_{\pm}=\pm \gamma_5 \Psi(y)_{\pm}$. Thus, $\bar\Psi_{\pm}\Psi_{\pm}$ 
is odd under $Z_2$, and the Dirac mass term, which is also odd, 
can be parametrized as $m_{\Psi}=c\sigma'$. The Dirac mass should 
therefore originate from the coupling to a $Z_2$ odd scalar field 
which acquires a vev. 
On the other hand, $\bar\Psi_{\pm}\Psi_{\mp}$ is even.
Using the metric (\ref{met}) one obtains for the left- and right-handed components
of the Dirac spinor \cite{GN,GP}
\begin{equation}
[e^{2\sigma}\partial_{\mu}\partial^{\mu}+\partial_5^2-\sigma'\partial_5-M^2]e^{-2\sigma}\Psi_{L,R}=0,
\end{equation} 
where $M^2=c(c\pm1)k^2\mp c\sigma''$ and  $\Psi_{L,R}=\pm\gamma_5\Psi_{L,R}$.

Decomposing the 5d fields as 
\begin{equation}
\Psi(x^{\mu},y)=\frac{1}{\sqrt{2\pi R}}\sum_{n=0}^{\infty}\Psi^{(n)}(x^{\mu})f_n(y),
\end{equation}
one ends up with a zero mode wave function \cite{GN,GP}
\begin{equation}
f_0(y)=\frac{e^{(2-c)\sigma}}{N_0},
\end{equation}
and a tower of KK excited states
\begin{equation}
f_n(y)=\frac{e^{5\sigma/2}}{N_n}\left[J_{\alpha}(\frac{m_n}{k}e^{\sigma})+
                 b_{\alpha}(m_n)Y_{\alpha}(\frac{m_n}{k}e^{\sigma})\right].
\end{equation}
The order of the Bessel functions is $\alpha=|c\pm 1/2|$ for $\Psi_{L,R}$.
The spectrum of KK masses $m_n$ and the coefficients $b_{\alpha}$
are determined by the boundary conditions of the wave functions at the 
branes.
The normalization constants follow from
\begin{equation}
\frac{1}{2\pi R}\int^{\pi R}_{\pi R}dy~e^{-3\sigma}f_m(y)f_n(y)=\delta_{mn}.
\end{equation}

Because of the orbifold symmetry, the zero mode of 
$\Psi_+$ $(\Psi_-)$ is a left-handed (right-handed) Weyl spinor.  
For $c>1/2$ $(c<1/2)$ the fermion is localized near the boundary
at $y=0$ $(y=\pi R)$, i.e.~at the Planck-  (TeV-) brane.

The zero modes of leptons and quarks acquire masses from their 
coupling to the Higgs field
\begin{equation} \label{3.1}
\int d^4x\int dy \sqrt{-g}\lambda^{(5)}_{ij}H \bar\Psi_{i+}\Psi_{j-}
\equiv \int d^4x ~ m_{ij} \bar\Psi_{iR}^{(0)}\Psi_{jL}^{(0)} +\cdots,
\end{equation}
where $\lambda^{(5)}_{ij}$ are the 5d Yukawa couplings. The 
4d Dirac masses are given by
\begin{equation} \label{3.2}
m_{ij}=\int_{-\pi R}^{\pi R}\frac{dy}{2\pi R}\lambda^{(5)}_{ij}e^{-4\sigma}H(y) f_{0iL}(y)f_{0jR}(y).
\end{equation}
Recall that the Higgs field is confined to the TeV-brane, 
i.e.~$H(y)=H_0\delta(y-\pi R)$
Using the known mass of the W-boson we can fix $H_0$ in terms of the 
5d weak gauge coupling $g^{(5)}$. 

The fermion wave functions and consequently their masses depend 
on the 5d mass parameters of the
left- and right-handed fermions, $c_L$ and $c_R$ respectively, 
which enter (\ref{3.2}).
As the 5d Dirac mass, i.e.~$c$ 
parameter of the fermion increases, the closer it gets closer localized 
towards the Planck-brane. Its overlap with the Higgs profile 
at the TeV-brane is consequently reduced, which is reflected in 
a smaller 4d fermion mass after electroweak symmetry breaking.
In ref.~\cite{HS2} we have shown that this geometrical picture 
beautifully generates the charged 
lepton and quark mass hierarchies, as well as quark mixings, 
by employing $c$-parameters and $\lambda_{ij}^{(5)}\sqrt{k}$ 
of order unity. In the following we apply this mechanism to 
also generate small neutrino masses consistent with neutrino
oscillation experiments.

\section{Neutrino masses}
In ref.~\cite{GN} it was have suggested that in 
warped geometry models small Dirac neutrino masses 
arise from a coupling to sterile (right-handed) bulk neutrinos. 
In order to generate masses in the sub-eV range, the sterile 
neutrinos have to be localized close to the Planck-brane, while 
the SM model neutrinos were confined to the 
TeV-brane. Here we generalize the scenario to incorporate 
bulk SM neutrinos.  

In our framework the SM fermions, described by 4d Weyl spinors,
correspond to left- and right-handed zero modes of 5d Dirac 
spinors which live in the bulk. The 5d mass parameters of the
$SU(2)$ doublet (singlet) leptons $L_i$ $(E_i)$ are $c_L^{(i)}$ 
$(c_{E}^{(i)})$, where $i=1,2,3$.  The right-handed neutrinos $\psi_i$
are also associated with bulk fermion fields which have 
right-handed zero modes. Their 5d mass parameters are denoted 
by $c_{\psi}^i$, where again $i=1,2,3$. We avoid large Majorana
neutrino masses by imposing lepton number as discussed earlier. 
After electroweak 
symmetry breaking the Dirac masses for the neutrinos are generated from 
the Yukawa-type coupling between the SM and sterile neutrinos, analogous
to eq.~(\ref{3.1})
\begin{equation}
{\cal L}=h^{(5)}_{ij}\bar{L}_i\psi_jH+{\rm h.c.~} +~\dots
\end{equation}
We will demonstrate that the neutrino oscillation data can be
reproduced with order unity Yukawa couplings $h^{(5)}_{ij}$.

From the KK reduction of the left-handed neutrino fields $\nu_L$
we obtain a left-handed zero mode $\nu_L^{(0)}$, corresponding to
the SM neutrinos,  and an infinite
tower of left- and right-handed KK excited states $\nu_L^{(i)}$ and 
 $\nu_R^{(i)}$, where we omit flavor indices. The sterile neutrinos
decompose into the right-handed zero mode $\psi_R^{(0)}$ and the
KK excited states $\psi_L^{(i)}$ and $\psi_R^{(i)}$. After electroweak
symmetry breaking the mass matrix takes the form
\begin{equation} \label{nu_mass}
M_{\nu}= (\nu_L^{(0)},\nu_L^{(1)}, \psi_L^{(1)},\dots) \left(\begin{array}{cccc} 
m^{(0,0)} & 0 &m^{(0,1)} & \cdots \\[.1cm] 
m^{(1,0)} & m_{L,1} & m^{(1,1)}& \cdots \\[.1cm]  
0 & 0 & m_{\psi,1} & \cdots \\
\vdots & \vdots & \vdots & \ddots
\end{array}\right)\left(\begin{array}{c}\psi_R^{(0)} \\[.1cm]   \nu_R^{(1)} \\[.1cm]   
\psi_R^{(1)} \\ \vdots \end{array}\right) 
\end{equation} 
where we again suppress flavor indices, i.e.~every entry represents
a $3\times 3$ matrix in flavor space. The masses $m^{(i,j)}$ 
are obtained by inserting the relevant wave functions
into eq.~(\ref{3.2}). The $m_{L,i}$ and $m_{\psi,i}$ are the KK masses
of the excited neutrino states. The zeros in (\ref{nu_mass}) follow 
from the $Z_2$ orbifold properties of the wave functions.
In ref.~\cite{HS} we used an exponential Higgs profile localized 
at the TeV-brane instead of strictly confining it to
the brane. In that case a mass term arises for $\psi_L^{(1)}\nu_R^{(1)}$, etc.
We have checked that including this term only affects the neutrino properties
at the $10^{-3}$ level (at most).

The squares of the physical neutrino masses are the eigenvalues of the 
hermitian mass matrix $M_{\nu}M_{\nu}^{\dagger}$. The unitary matrix $U$,
such that $U^{\dagger}M_{\nu}M_{\nu}^{\dagger}U$ is diagonal,
relates the left-handed mass eigenstates $N_L^{\rm phys}$ to the 
interaction eigenstates $N_L=(\nu_L^{(0)},\nu_L^{(i)},\psi_L^{(i)})$ via
$N_L=UN_L^{\rm phys}$. The physical neutrinos $\nu_L^{\rm phys}$ 
correspond to the three lightest states in $N_L^{\rm phys}$. The 
right-handed mass eigenstates are obtained from a unitary matrix
$V$ which diagonalizes  $M_{\nu}^{\dagger}M_{\nu}$.
If the Yukawa couplings of the charged leptons are diagonal, the
mixing angles of the physical neutrinos can be directly read off 
from $U$. 

The atmospheric and solar neutrino anomalies can be solved
by assuming oscillations among the different neutrino flavors
via the mass matrix $M_{\nu}$ (see e.g.~\cite{osc} for 
a recent review). The atmospheric neutrino flux is reproduced
by $\Delta m^2_{\rm atm}=2$--$7\cdot10^{-3}$ eV$^2$
and an almost maximal mixing angle $\sin^22\theta_{\rm atm}\sim1$. 
For the solar neutrinos a variety of masses and mixing
angles fit the data. For the large mixing angle MSW solution one finds
$\Delta m^2_{\rm sol}=10^{-5}$--$10^{-3}$ eV$^2$
and $\sin^22\theta_{\rm sol}\sim1$, while for the small mixing angle 
MSW solution $\Delta m^2_{\rm sol}=5\cdot10^{-6}$--$10^{-5}$ eV$^2$
and $\sin^22\theta_{\rm sol}=10^{-3}$--$10^{-2}$. Similarly, for 
the LOW solution $\Delta m^2_{\rm sol}=5\cdot10^{-8}$--$10^{-7}$ eV$^2$
and $\sin^22\theta_{\rm sol}\sim1$ and for the vacuum 
solution $\Delta m^2_{\rm sol}\sim 10^{-10}$ eV$^2$
and $\sin^22\theta_{\rm sol}\sim1$. Finally, the CHOOZ
reactor experiment together with the atmospheric
neutrino data constrain $|U_{e3}|^2$ to be smaller
than a few times $10^{-2}$. In the next section
we demonstrate how neutrino mass matrices with
these properties can be obtained from bulk neutrinos. 
 
Our approach to neutrino oscillations is in the spirit of
``neutrino mass anarchy'' models \cite{HMW00}. It has been
demonstrated that a fair fraction of neutrino mass matrices, 
which appear to have random entries, are consistent with the
experimental constraints. This works best for the large MSW
solution to the solar neutrino anomaly, where the ratio of 
$\Delta m_{\rm atm}^2/\Delta m_{\rm sol}^2$ is not too large, and no tiny 
mixing angles are involved. In our analysis of neutrino
oscillations we assume non-hierarchical Yukawa
couplings $h_{ij}^{(5)}$ on order of the 5d gauge coupling $g_2^{(5)}$. 
However, in our framework the structure of the Yukawa couplings
is not mapped one to one to a neutrino mass matrix. Rather, the
Yukawa texture is deformed by the neutrino wave function
in the extra dimension which enter eq.~(\ref{3.2}). Therefore,
it is not trivial that the neutrino data can be obtained with
order unity Yukawa couplings. On the other hand, one may
hope that the wave functions induce hierarchies
that allows one to also reproduce also the small angle MSW, 
LOW and vacuum solution to the solar neutrino problem.

\section{Numerical results}
Various constraints on the scenario with bulk gauge and fermion fields
have been discussed in the literature \cite{CHNOY,GP,HS,DHR,DHR2,dAS}.
With bulk gauge fields for instance, the SM relationship between the gauge couplings 
and masses of the Z and W bosons gets modified. The electroweak 
precision data then requires the lowest KK excitation of the gauge bosons 
to be heavier than about 10 TeV \cite{HS}.
This bound becomes especially important if the fermions are localized
towards the Planck-brane $(c>1/2)$. As discussed in ref.~\cite{HS2}, 
this applies to all SM fermions, with the exception of the top-quark, if the 
fermion mass hierarchy arises from their different locations in the
extra dimension. In this case the bounds induced by the contribution
of KK  excitations of the SM gauge bosons to  the electroweak 
precision observables are weak. They only require the KK masses
to be above about 1 TeV \cite{GP,DHR}. 
In the following we will therefore assume that
the mass of the first KK gauge boson is $m_1^{(G)}=10$ TeV.
The corresponding masses of the KK fermions are then in the range
10 to 16 TeV, for $0<c<1.5$. The mass of the lightest KK graviton
is 16 TeV.

For simplicity we assume diagonal Yukawa couplings $\lambda^{(5)}_{ij}$ 
for the charged leptons. Neutrino mixing is then solely governed by
the neutrino mass matrix $M_{\nu}$. To avoid a hierarchy in the 5d couplings, 
we assume $\lambda^{(5)}_{ii}=g_2^{(5)}$, where the 5d weak gauge
coupling $g_2^{(5)}\sim g_2\sqrt{2\pi R}$. 
We take $k=M_5=\overline M_{\rm Pl}$, where $\overline M_{\rm Pl}=2.44\times10^{18}$
GeV is the reduced Planck mass. From $m_1^{(G)}=10$ TeV 
we determine the brane separation $kR=10.83$ \cite{DHRP}.

\begin{figure}[t] 
\begin{picture}(100,160)
\put(95,-10){\epsfxsize7cm \epsffile{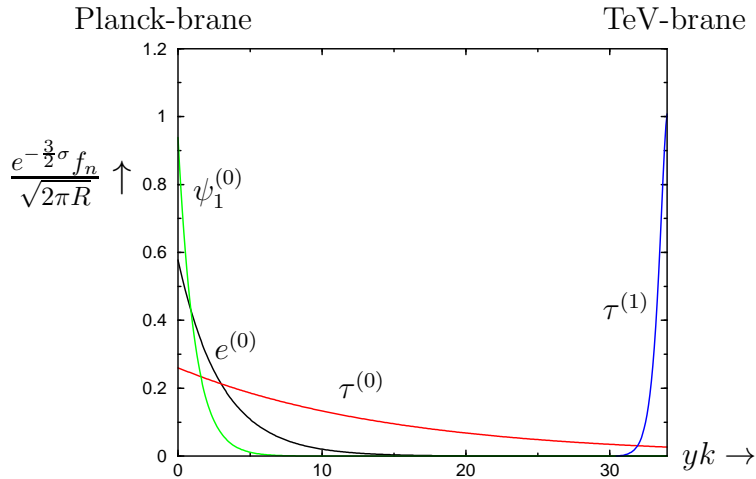}}
\put(45,100){{\large $\frac{e^{-\frac{3}{2}\sigma}f_n}{\sqrt{2\pi R}}\uparrow$}} 
\put(115,95){{$\psi_1^{(0)}$}} 
\put(170,20){{$\tau^{(0)}$}} 
\put(123,35){{$e^{(0)}$}} 
\put(270,50){{$\tau^{(1)}$}} 
\put(300,-5){$yk\rightarrow$}
\put(270,160){{TeV-brane}} 
\put(70,160){{Planck-brane}} 
\end{picture} 
\caption{Localization of the electron, $\tau$ and $\psi_1$ zero modes,
and the first KK state of the $\tau$ in the extra dimension for the 
parameters of eq.~(\ref{SMSW}).
}
\label{f_1}
\end{figure}

As discussed in ref.~\cite{HS2} the measured charged lepton
masses do not determine the 5d mass parameters  $c_L^{(i)}$ 
and $c_{E}^{(i)}$ in a unique way. In order to minimize Majorana
neutrino masses, we proposed a scenario where the left-handed leptons $L_i$
were placed as close as possible towards the Planck-brane, while
the right-handed leptons $E_i$ were assumed to be delocalized. The associated 
fermion mass parameters are \cite{HS2}
\begin{equation}
 c_L^{(1)}=0.834,~ c_L^{(2)}=0.664, ~c_L^{(3)}=0.567,~~~ 
c_E^{(1)}= c_E^{(2)}= c_E^{(3)}=0.5~.
\end{equation}
The zero mode wave functions of the left-handed electron and tau,
the right-handed zero mode of $\psi_1$ and the first excited
state of the tau are displayed in fig.~\ref{f_1}. 
Since the SM neutrinos are at different locations in the
extra dimension, non-hierarchical Yukawa couplings
typically lead to nearest neighbor-type neutrino
mass matrices.  This is similar to the case of quark
mixings which we discussed in ref.~\cite{HS2}. 
By taking the Yukawa coupling of $\nu_{\tau}$ to be 
somewhat smaller than unity, it is however possible
to obtain large  $\nu_{\mu}$--$\nu_{\tau}$  mixing.
Large $\nu_{e}$--$\nu_{\mu}$  mixing would require 
a hierarchy of order $10^2$ at least in the Yukawa 
couplings, which leaves the small MSW scenario as
the only natural solution to the solar neutrino anomaly. 
The data are, for instance, reproduced by taking
\begin{eqnarray} \label{SMSW}
c_{\psi}^{(1)}=1.38,~ c_{\psi}^{(2)}=1.29, ~c_{\psi}^{(3)}=1.24, 
\nonumber\\[.2cm]
\frac{h_{ij}^{(5)}}{g_2^{(5)}}=\left(\begin{array}{ccc} 
1.1 & 2.9 & 3.3 \\
-2.1 & -3.6 & 3.8 \\
-0.5 & -0.4 & 0.3
\end{array}\right).
\end{eqnarray}
For the light neutrino masses we obtain $m_{\nu1}=8.0\cdot 10^{-6}$ eV,
$m_{\nu2}=2.6\cdot 10^{-3}$ eV and $m_{\nu3}=7.6\cdot 10^{-2}$ eV. The
lightest KK state is basically a $\nu_{\tau}$ excitation with a mass
of 10.4 TeV. The excitations of the other bulk fields have similar masses.
The neutrino data are successfully reproduced with $\Delta m^2_{\rm atm}=
5.8\cdot10^{-3}$ eV$^2$, $\sin^22\theta_{\rm atm}=0.92$,
$\Delta m^2_{\rm sol}=6.8\cdot10^{-6}$ eV$^2$ and 
$\sin^22\theta_{\rm sol}=3.0\cdot10^{-3}$. The quantity $U_{e3}^2=5\cdot 10^{-6}$, 
much smaller than the experimental bound. This example 
demonstrates that the small angle MSW solution can be obtained from 
order unity parameters. This is in contrast to ``neutrino mass anarchy 
models'' in four dimensions \cite{HMW00}. 

The SM neutrinos mix not only with each other but also
with the left-handed KK states of the sterile neutrinos 
$\psi_L^{(i)}$ \cite{GN}. This effect diminishes the effective weak
charge of the light neutrinos. As a result the effective number
of neutrinos contributing to the width of the Z boson is reduced
to $n_{\rm eff}=3-\delta n$, where $\delta n$ is obtained from 
summing the relevant squared entries of $U$. Measurements
of the Z width induce the constraint $\delta n \lsim 0.005$ \cite{zwidth}.
For the parameter set (\ref{SMSW}) we find the result $\delta n = 6\cdot 10^{-7}$, 
well below the experimental sensitivity.
Because of this small mixing with KK states, the masses and mixings
of the light neutrinos can be reliably computed from the zero mode
part of the neutrino mass matrix (\ref{nu_mass}). In ref.~\cite{GN}
a result three orders of magnitude larger was reported. Our result is
smaller since our KK masses are $\sim$10 TeV instead of 1 TeV, and
being bulk fields, the SM
neutrinos have a smaller overlap with the Higgs field at the TeV-brane
which reduces the off-diagonal entries $m^{(0,i)}$ in  (\ref{nu_mass}).
 
In order to reproduce the large mixing angle solutions to the solar
neutrino problem we have to relocate the left-handed lepton fields
in the extra dimension. A nearest neighbor-type neutrino mass matrix
is avoided by placing the SM neutrinos all at the same position, 
i.e.~$c_L^{(1)}=c_L^{(2)}=c_L^{(3)}\equiv c_L$. We choose $c_L=0.567$.
In this case our results concerning proton decay \cite{HS2} are 
still valid. The charged fermion masses are reproduced with
$c_E^{(1)}=0.787$, $c_E^{(2)}=0.614$ and $c_E^{(3)}=0.50$. 
A parameter set which implements the large angle MSW solution is
\begin{eqnarray} \label{LMSW}
c_{\psi}^{(1)}=1.43,~ c_{\psi}^{(2)}=1.36, ~c_{\psi}^{(3)}=1.30, 
\nonumber\\[.2cm]
\frac{h_{ij}^{(5)}}{g_2^{(5)}}=\left(\begin{array}{ccc} 
-2.0 & 1.5 & -0.5 \\
-1.8 & -1.1 & 1.9 \\
0.5 & 1.9 & 1.7
\end{array}\right).
\end{eqnarray}
We obtain the light neutrino masses $m_{\nu1}=1.0\cdot 10^{-3}$ eV,
$m_{\nu2}=1.0\cdot 10^{-2}$ eV and $m_{\nu3}=7.1\cdot 10^{-2}$ eV.
The KK spectrum is similar to the small MSW case.  
For the neutrino oscillation parameters we find $\Delta m^2_{\rm atm}=
4.9\cdot10^{-3}$ eV$^2$, $\sin^22\theta_{\rm atm}=0.99$,
$\Delta m^2_{\rm sol}=1.0\cdot10^{-4}$ eV$^2$ and 
$\sin^22\theta_{\rm sol}=0.90$. Also, $U_{e3}^2=0.036$ 
is close to the experimental bound which is typical for the 
solutions we find. Since the SM neutrinos now are closer 
to the TeV-brane, their mixing with the sterile neutrinos is
enhanced. We find $\delta n =2\cdot 10^{-5}$, still well below
the experimental bound. 

Along the same lines it is also possible to reproduce the LOW
and the vacuum solution to the solar neutrino anomaly. A smaller
$\Delta m^2_{\rm sol}$ is obtained by moving the sterile
neutrinos $\psi_1$ and $\psi_2$ closer towards the Planck-brane.
We find viable solutions with $c_{\psi}^{(1)}=1.50$ and  $c_{\psi}^{(1)}=1.45$
(LOW solution), and $c_{\psi}^{(1)}=1.62$ and  $c_{\psi}^{(1)}=1.57$
(vacuum solution). The mixing between SM neutrinos and sterile
neutrinos is of the size found in the large MSW solution.

\section{Rare processes}
The direct experimental signature for bulk SM fields would be
the production of the associated KK states in high energy colliders.
Since typical KK masses in our scenario are of order 10 TeV, this
is not possible before the advent of LHC.  It is therefore important to
study the implications of the excited states on rare processes and
precision variables. 

In order to prevent large Majorana neutrino masses, we imposed
lepton number. As a result the proton is stabilized as well. However,
baryon number may still be violated by $\Delta B=2$ operators 
contributing to neutron anti-neutron oscillations and double 
nucleon decay \cite{BD,CC01}. To be specific let us consider
the 6-fermion operator $O_{\Delta B=2}=U_1D_1D_2U_1D_1D_2$.
The right-handed quark fields $U_i$ and $D_i$ are assumed to
be bulk fields. Experiments constrain the suppression scale of this
operator by $M_{\Delta B=2}\gsim 3.3\cdot 10^5$ GeV \cite{CC01}.
In ref.~\cite{HS2} we discussed how the quark masses and mixings
can be obtained by locating them at different positions in the
extra dimension. Using the 5d quark mass parameters from that
analysis,  $c_{D1}=0.57$, $c_{U1}=0.63$, $c_{D2}=0.57$, we 
estimate  $M_{\Delta B=2}\sim 8\cdot 10^5$ GeV.  The computation
of $M_{\Delta B=2}$ is a straight forward generalization of the 
4-fermion operator case discussed in refs.~\cite{GP,HS2}, where
the effective 4d operator is obtained by integrating the fermion wave 
functions over the extra dimension. 
Since in 4 dimensions $O_{\Delta B=2}$ is suppressed by five
powers of   $M_{\Delta B=2}$, the rate of $\Delta B=2$
processes is about $10^{-2}$ below the experimental bound. 
However, the 5d mass parameters are not uniquely fixed by
their masses and mixings. If we move the quark fields closer to
the TeV-brane, e.g.~ $c_{D1}=0.54$, $c_{U1}=0.60$, $c_{D2}=0.54$,
 $M_{\Delta B=2}$ comes down to the experimental limit. The quark
masses and mixings are recovered by reducing the quark Yukawa
coupling by a common factor of about 5 compared to the values given
in ref.~\cite{HS2}. Flavor changing operators are still sufficiently 
suppressed by a scale of about $10^6$ GeV.  

Non-zero neutrino masses violate the lepton flavor symmetry, which
induces processes like $\mu\rightarrow e\gamma$. The rate for these
transitions is considerable enhanced by the 
presence of heavy neutrino states \cite{CL77,IP94}
\begin{equation} \label{meg}
\Gamma (\mu\rightarrow e\gamma)\propto \left|\sum_i U_{ei}^*U_{\mu i} ~{\cal F}
                    \left(\frac{m_i^2}{M_W^2}\right)\right|^2.
\end{equation}
The matrix elements $U_{ei}$ and $U_{\mu i}$ encode the 
admixture of the $i$th mass eigenstate in $\nu_{e}$ and
$\nu_{\mu}$, respectively. If all neutrino masses are much 
smaller than $M_W$, i.e.~${\cal F}(m_i^2/M_W^2)\sim{\cal F}(0)$, 
the rate (\ref{meg}) is suppressed due to the unitarity of $U$. 
Heavy neutrinos prevent this cancellation. Lepton flavor violation
in the warped geometry model with TeV-brane SM neutrinos and
bulk sterile neutrinos \cite{GN} was studied in ref.~\cite{K00}.
It was found that the branching ratio for  $\mu\rightarrow e\gamma$
exceeds the experimental limit, 
Br$(\mu\rightarrow e\gamma)<1.2\cdot10^{-11}$ \cite{B00}, 
if the first KK state of the sterile neutrino has a mass below 25 TeV. 
Thus, lepton flavor violation imposes the most stringent constraint on 
the model. 

Evaluating the branching ratio from (\ref{meg}) in our
scenario with bulk SM fermions  leads to a somewhat different 
conclusion. The rate for $\mu\rightarrow e\gamma$ is very
sensitive to the mixing between light and heavy neutrino states.
With bulk neutrinos the mixing with heavy states is considerably
reduced, as was discussed in the previous section.
Therefore, we expect that the constraints from lepton flavor violation
should be less stringent in our scenario.  Indeed, for 
the example implementing the small MSW solution (\ref{SMSW}), we find 
Br$(\mu\rightarrow e\gamma)\sim 10^{-26}$, which is far below the
experimental sensitivity. In the case of the large  MSW solution (\ref{SMSW})
the result is considerably enhanced because of the larger mixing of
SM and sterile neutrinos. We obtain Br$(\mu\rightarrow e\gamma)\sim 10^{-15}$,
still well below the experimental sensitivity. The branching ratios for 
$\tau\rightarrow e\gamma$ and $\tau\rightarrow \mu\gamma$
are also much too small to be observed in present 
experiments. For the LOW and vacuum solutions to
the solar neutrino anomaly, the rate for lepton flavor violation is similar
to the result for the large MSW example. 

In principle, the rates for lepton flavor violation can
be considerably enhanced by moving the SM neutrinos (and with 
them the charged left-handed leptons) closer to the TeV-brane. 
For instance, if we take $c_L=0.5$ as the common 5d mass parameter 
of the left-handed leptons, we can obtain Br$(\mu\rightarrow e\gamma)\sim 10^{-13}$
with the large MSW solution. The current experimental sensitivity is 
almost reached for $c_L=0.45$ which leads to  
Br$(\mu\rightarrow e\gamma)\sim 5\cdot 10^{-12}$. In order to reproduce
the charged lepton and neutrino masses, the right-handed charged
leptons and sterile neutrinos have to reside closer to the Planck-brane.
However, for $c_L<0.5$ the effective 4d weak gauge of the left-handed 
leptons start to deviate from the gauge couplings of the right-handed
leptons that live close to the Planck-brane, since the W and Z boson
wave functions are not strictly constant in the extra dimension \cite{HS}. 
The associated violation of lepton universality renders this part of 
the parameter space less attractive. For $c_L>0.5$, lepton 
flavor violation may still be only a few orders of magnitude below
the present experimental limit.

\section{Conclusions}
We have shown how atmospheric and solar neutrino oscillations, especially
the bimaximal mixing scenario, can be incorporated in models with a small
extra dimension and non- factorizable ( warped) geometry. An important
distinction from earlier work \cite{GN} is the appearance of the SM
fermions in our case as bulk fields. Rare processes such as $n- \bar n$
oscillations and $\mu\rightarrow e+\gamma$ may occur at a rate 
not too far below the current experimental limits.

\section*{Acknowledgement}
S.~H.~is supported in part by the Feodor Lynen Program of the
Alexander von Humboldt foundation. This work was also supported 
by DOE under contract DE-FG02-91ER40626.

\end{document}